\documentclass[preprint]{aastex}

\shorttitle{Close binary stars.~XIII}
\shortauthors{Rucinski \& et al.}

\begin{document}

\title{Radial Velocity Studies of Close Binary
Stars.~XIII\footnote{Based on the data obtained at the David Dunlap
Observatory, University of Toronto.}}

\author{}
\author{
Slavek M. Rucinski, 
Theodor Pribulla\altaffilmark{2},
Stefan W. Mochnacki, 
Evgenij Liokumovich,\\ 
Wenxian Lu\altaffilmark{3},
Heide DeBond, Archie de Ridder, Toomas Karmo, Matt Rock, J.R. Thomson}
\affil{David Dunlap Observatory, University of Toronto \\
P.O.~Box 360, Richmond Hill, Ontario, Canada L4C~4Y6}
\email{(rucinski,mochnacki,debond,ridder,karmo)@astro.utoronto.ca,pribulla@ta3.sk}
\author{Waldemar Og{\l}oza}
\affil{Mt. Suhora Observatory of the Pedagogical University\\
ul.~Podchora\.{z}ych 2, 30--084 Cracow, Poland}
\email{ogloza@ap.krakow.pl}

\author{Krysztof Kaminski, Piotr Ligeza}
\affil{Adam Mickiewicz University Observatory, S{\l}oneczna 36, 
60--286 Pozna\'{n}, Poland}
\email{chrisk@amu.edu.pl,piotrl@amu.edu.pl}

\altaffiltext{2}
{Astronomical Institute, Slovak Academy of Sciences, 
059 60 Tatransk\'a Lomnica, Slovakia}

\altaffiltext{3}
{Current address: Department of Geography and Meteorology and
 Department of Physics and Astronomy, Valparaiso University, Valparaiso,
 IN 46383, U.S.A. e-mail: Wen.Lu@valpo.edu}

\begin{abstract}
Radial-velocity measurements and sine-curve fits to the orbital radial
velocity variations are presented for ten close binary systems: EG~Cep,
V1191~Cyg, V1003~Her, BD+7\degr3142, V357~Peg, 
V407~Peg, V1123~Tau, V1128~Tau, HH~UMa, and PY~Vir.
While most of the studied eclipsing systems are contact binaries, EG~Cep is
a detached or a semi-detached double-lined binary and V1003~Her is a close
binary of an uncertain type seen at a very low inclination 
angle. We discovered two previously unknown triple systems,
BD+7\degr3142 and PY~Vir, both with late spectral-type (K2V) binaries. 
Of interest is the low-mass ratio ($q = 0.106$) 
close binary V1191~Cyg showing an extremely fast period increase; 
the system has a very short period for its spectral type and shows a
W-type light curve, a feature rather unexpected for such a 
low mass-ratio system. 
\end{abstract}

\keywords{ stars: close binaries - stars: eclipsing binaries --
stars: variable stars}

\section{INTRODUCTION}
\label{sec1}

This paper is a continuation of the series of papers (Papers
I -- XII) of radial-velocity studies of close binary stars
and presents data for the twelfth group of ten close binary stars observed
at the David Dunlap Observatory. For full references to the
previous papers, see the last paper by \citet[ Paper XII]{ddo12};
for technical details and conventions, for preliminary
estimates of uncertainties, and for a description of the
broadening functions (BFs) technique, see the interim summary
paper by \citet[ Paper VII]{ddo7}. The DDO studies
use the efficient program of \citet{pych2004}
for removal of cosmic rays from 2-D images.

All data used in the present paper were obtained using the broadening
functions extracted from the region of the Mg~I triplet at
5184~\AA, as in most of the previous papers. In August 2005, a new 
2160 lines/mm grating was acquired to replace the 
previously most frequently used 1800 lines/mm
grating which after many years of use lost its efficiency. 
The new grating markedly improved quality of the observed 
spectra and of the resulting BFs. Of the reported results, 
the older grating was used for 1997 observations 
of V357~Peg, V1123~Tau and V1128~Tau (this star was also briefly
observed in 2003); all three stars were later re-observed 
using the new grating. The radial velocity (hereafter RV) observations 
reported in this paper have been collected between October 1997 and August 2007. 
The ranges of dates for individual systems can be found in Table~\ref{tab1}.
A few additional, low quality spectra of PY~Vir, not listed in Table~\ref{tab1},
were taken on February 19, 2008.

Selection of the targets in our program remains quasi-random: At a given time,
we observe a few dozen close binary systems with periods usually
shorter than one day, brighter than 10 -- 11 magnitude and with
declinations $>-20^\circ$; we publish the results in groups of ten systems
as soon as reasonable orbital elements are obtained from measurements evenly
distributed in orbital phase. For this paper, we selected mostly 
targets from among fainter {\it HIPPARCOS\/} discoveries.
None of the present targets has a spectroscopic orbit published. 
V1003~Her and PY~Vir have only relatively poor, ground-based 
photometric data while 
V1191~Cyg is a rather neglected but -- as we describe -- a 
very interesting contact binary.

The radial velocities for the short period binaries reported 
in this paper were determined by fitting the double rotational profiles, 
as explained in \citet{ddo11}.
Similarly as in our previous papers dealing with multiple systems
(here the cases of BD+7\degr3142 and PY~Vir), RV's for the eclipsing pair
were obtained after removal of the slowly rotating components
\citep{ddo12}. 

As in other papers of this series, whenever possible, we estimate spectral
types of the program stars using new classification spectra centered
at 4200 \AA\ or 4400 \AA. 
These are compared with the mean $(B-V)$ color indices usually
taken from the Tycho-2 catalog \citep{Tycho2} and the photometric estimates
of the spectral types using the relations of \citet{Bessell1979}.
In this paper we also made use of infrared colors determined
from 2$\mu$ All Sky Survey \citep[ 2MASS]{2MASS}. Especially useful is the
$J-K$ color index, which is monotonically rising from the early spectral types to
about M0V \citep{cox2000}. This infrared color is affected relatively
less by the interstellar absorption than $B-V$.
Parallaxes cited throughout the paper were adopted from the new reduction
of the {\it HIPPARCOS\/} data \citep{newhipp} which supersede the original
reductions \citep{hip}.

This paper is structured in a way similar to that of previous papers, in that
most of the data for the observed binaries are in two tables consisting of the
RV measurements in Table~\ref{tab1} and of their preliminary sine-curve solutions
in Table~\ref{tab2}. Radial velocities 
and the corresponding spectroscopic orbits for all
ten systems are shown in phase diagrams in Figures~\ref{fig1} -- \ref{fig3}.
The measured RV's are listed in Table~\ref{tab1}. Only data with full weights
were used to compute the orbits; we discarded poorer observations.
Table~\ref{tab2} contains 
also our new spectral classifications of the program objects. Section~\ref{sec2} 
of the paper contains summaries of previous studies for individual systems 
and comments on the new data. Examples of BFs of individual systems extracted 
from spectra observed close to quadratures are shown in Fig.~\ref{fig4}.

The data in Table~\ref{tab2} are organized in the same manner as in
the previous papers of this series. 
In addition to the parameters of spectroscopic orbits,
the table provides information about the relation between the
spectroscopically observed upper conjunction of the more massive
component, $T_0$ (not necessarily the primary eclipse)
and the recent photometric determinations of the primary
minimum in the form of the $O-C$ deviations for the number of
elapsed periods $E$. The reference ephemerides were taken from
various sources: 
For V1003~Her, we doubled the {\it HIPPARCOS\/} period and shifted the instant
of the maximum by $0.25P$; for BD+7\degr3142, we took it from the 
ASAS\footnote{http://archive.princeton.edu/$\sim$asas/}
survey \citep{asas}; for V407~Peg, from \citet{maciej2002} 
and for PY~Vir, from \citet{wils2003}. 
For the rest of the systems, the ephemerides
given in the on-line version of ``An Atlas O-C diagrams of eclipsing binary
stars''\footnote{http://www.as.wsp.krakow.pl/ephem/} \citep{Kreiner2004}
were adopted. Because the on-line ephemerides are frequently updated, we
give those used for the computation of the $O-C$ residuals in the comments to
Table~\ref{tab2} (the status as of December 2007). The deeper eclipse in W-type
contact binary systems corresponds to the lower conjunction of the more massive
component; in such cases the epoch in Table~\ref{tab2} is a half-integer number.

\section{RESULTS FOR INDIVIDUAL SYSTEMS}
\label{sec2}

\subsection{EG~Cep}

The variability of EG~Cep was discovered by \citet{stroh1958}. 
Photoelectric light curves and their solutions have been presented by 
several investigators; for references see \citet{erdem05}.
The system shows a $\beta$~Lyrae-type light curve 
with minima about 1.0 and 0.3 mag deep.
\citet{etzel1993} determined the projected rotational velocity of 
the primary as $146 \pm 20$ km~s$^{-1}$. A
photometric analysis of the system was performed 
by \citet{kaluz1984} who used a grid search to find the 
mass ratio $q_{ph} = 0.45 - 0.50$ and a semi-detached 
configuration for the system. 
\citet{choch1998} discussed the long-term period change in the 
system and also arrived at the semi-detached configuration 
with the less massive component filling its 
Roche lobe and the mass ratio $q_{ph}$ = 0.47. 
The new analysis of \citet{erdem05} led to a similar 
mass ratio. No spectroscopic orbit of the system has been published 
so far.

In view of discrepancies between
the spectroscopic and photometric mass ratios previously noted in
our DDO series, it is surprising, but also encouraging, that 
all photometric investigations have led to a mass ratio close 
to our spectroscopic value, $q_{sp}$ = 0.464(5). 
This is in spite of the semi-detached configuration which 
has weaker photometric constraints than the contact configuration
in the light curve solution. The agreement
must result from the high orbital inclination angle, 
$\approx 86\degr$ \citep{choch1998} and the presence of the total
eclipses.

The intrinsic color of the system was estimated at about 
$(B-V)_0$ = 0.197 \citep{kaluz1984} indicating the spectral type 
of the primary component of A7. 
The Tycho-2 color index $B-V=0.24$ is consistent with our
A7V classification and requires a slight reddening.
The $J-K = 0.19(4)$ color indicates the A9V -- F2V spectral type
which may reflect the stronger contribution of the
secondary component. EG~Cep was not included into the {\it HIPPARCOS\/}
astrometric measurements.

\subsection{V1191~Cyg}

The variability of the contact binary V1191~Cyg (GSC 3159-1512, 
$V_{max}$ = 10.82, $V_{min}$ = 11.15) was detected
by \citet{mayer1965} while observing the nearby star V1187~Cyg. 
Since then the system was rather neglected with 
the only thorough photometric analysis of  
\citet{prib2005}, who found that the orbital 
period of the system increases at a record rapid
(for contact binaries) 
rate of $\Delta P/P$ = 2.12~10$^{-6}$ year$^{-1}$.  
The system is totally eclipsing so the geometric elements derived
through a light curve solution,
$i$ = 80.4\degr, $q$ = 0.094 and $f$ = 0.46 were well defined
in the solution of \citet{prib2005}. 
The deeper minimum is flat, hence the
system was classified as a W-type contact binary. 
The authors determined $(B-V) = 0.62$ (uncorrected for
the interstellar absorption). No spectroscopic study of the system 
has been published yet.

We found V1191~Cyg to be a rather 
difficult spectroscopic target due to its relative faintness, the
short orbital period of $P = 0.3134$ days and the weakness of
the Mg~I triplet lines. 
As a solution, we took 129 spectra evenly covering 
all phases; the extracted BFs were subsequently smoothed in 
the phase domain (the phase step of 0.02) in the way 
described before in \citet{ddo10}.

The spectroscopic lower conjunction of the 
more massive component occurs during the primary
minimum, hence V1191~Cyg is a W-type system. 
The spectroscopic mass ratio, $q_{sp} = 0.107 \pm 0.005$ 
is consistent with the photometric determination. The 
system is, however, very unusual: (1)~It is of a rather late 
spectral type for such a small mass ratio; (2)~Its mass ratio
is unusually small for a W-type system; 
(3)~The orbital period is short for the mid-F spectral type
implying smaller, more compact components than for typical
solar-type contact binaries. 
The 2MASS color, $J-K = 0.318$, corresponds to the
F6V spectral type which is exactly what we see in the
classification spectra. The Tycho-2 color, $B-V = 0.39 \pm 0.10$ 
is too uncertain to draw any firm conclusions.

\subsection{V1003 Her}

The variability of this system was found by the {\it HIPPARCOS\/} 
satellite. It is described there as a periodic variable of 
an unspecified variability type with the period $P = 0.246661$ days.
Later \citet{duer1997}, suggested that it is a 
contact binary with the orbital period twice the original 
{\it HIPPARCOS\/} value. Analysis of photometric observations of the binary
is complicated by the rather low amplitude, $\Delta V = 0.09$. 
In the ASAS-3 survey  \citep{asas}, 
the system appears with an undefined variability type and
the best period of 21.846 days; however, the same observations,
when phased with the double of the {\it HIPPARCOS\/} period, 
show a light curve with the
minima of similar depths, a feature which is typical for contact binaries.
Rather noisy observations can be found in the NSVS database (NSVS 11074663, 
http://skydot.lanl.gov/nsvs/nsvs.php). Except for those fragmentary 
observations, no other photometric or 
spectroscopic observations of V1003~Her are available. 

Our spectroscopic observations show that V1003~Her is indeed a close binary,
most likely of the W~UMa type, but -- because of the very small
photometric amplitude -- a full 
description and classification of the system is impossible at this
point and would require a high-precision photometry.
The projected total mass of the system, 
$(M_1+M_2) \sin^3 i = 0.672 \pm 0.009$ M$_\odot$, the low photometric  
amplitude, the early spectral type (A7) and the 
absence of any third light, all indicate a very low 
inclination angle. The {\it HIPPARCOS\/} parallax, 
$\pi = 4.64 \pm 1.66$ mas is of limited use because of its low
relative accuracy. Our spectral type estimate, A7V, is 
consistent with the 2MASS infrared color, $J-K$ = 0.234, 
but the Tycho-2 color index 
$B-V=0.38$ indicates a substantial reddening in the
direction to V1003~Her, $E(B-V) \approx 0.19$.

\subsection{BD+7\degr3142 (Her)}

Variability of BD+7$\degr$3142 was detected during the 
analysis of the ASAS data \citep{asas}.
It was classified as an eclipsing binary with the 
following ephemeris: 
$HJD = 2\,452\,383.92 + 0.275277 \times E$ for the primary
minima. No photometric or spectroscopic 
observations of the system have been published yet. 
Simultaneous photometry during our spectroscopic observations has
led to determination of one light minimum at $HJD = 2\,454\,188.8663(1)$, 
which occurred at the phase 0.8366 of the ASAS prediction. Our 
spectroscopic conjunction agrees with this newly observed minimum.

The spectroscopic observations show BD+7\degr3142 as 
a triple system with a third component stationary in radial velocities. 
The light contribution of the third 
component is $L_3/(L_1+L_2) = 0.50$ at the brightness maximum
of the eclipsing pair; its signature is narrow with 
$V \sin i \leq 15$ km~s$^{-1}$, which is close to the resolution
of our spectroscopy. After approximating the third peak by 
a Gaussian and its removal from the BFs, we obtained 
well-defined BFs of the eclipsing pair. 
The radial velocity of the third component,
$RV_3 = -69.2 \pm 1.4$ km~s$^{-1}$ was found to be close to the 
center of mass velocity of the eclipsing pair, 
$V_0 = -64.1 \pm 0.8$ km~s$^{-1}$, hence the third component 
appears to be a physical member of the system.  
The small difference of velocities is, very probably, a 
result of a slow mutual revolution. 
The system is not listed in the WDS catalogue as a previously
recognized visual double \citep{wds}.

The parallax of the system is unknown, but its absolute visual magnitude can be 
estimated using calibration of \citet{rd1997} and the K2 spectral type 
($(B-V)_0 = 0.74$) at $M_V = 4.84$. A correction of the maximum visual magnitude
of the combined system, $V = 9.89$, for the contribution of the third component,
results in $V_{12} = 10.33$ and then the distance $d = 97$ pc. 

By combining the proper motion of BD+7\degr3142
adopted from the Tycho-2 catalogue \citep{Tycho2} with 
the systemic radial velocity and the estimated distance, 
one obtains a large space velocity of 88 km~s$^{-1}$. 
The 2MASS color of BD+7$\degr$3142,
$J-K = 0.591$, and the Tycho-2 $B-V = 0.92$ agree with our
spectral type of K2V and is consistent with the 
very short period of the system.

\subsection{V357~Peg}

The variability of V357~Peg (HD222994) was discovered during the {\it HIPPARCOS\/}
mission. The system was correctly classified as a contact binary. 
The first photometry of V357~Peg, published
by \citet{yasar2000}, shows a typical W~UMa-type light curve. 
The photometric amplitude is about 0.48 mag so the eclipses 
cannot be far from total (assuming our $q_{sp} = 0.40$). 
No photometric or spectroscopic study of the system has been 
published yet in spite of a fairly large brightness of $V_{max} = 9.06$. 

Our spectroscopic observations show that 
V357~Peg is a contact binary of the A type with the 
mass ratio of $q = 0.401$ which is moderately large for this type. 
The total projected
mass of the system $(M_1 + M_2) \sin^3 i = 2.112\,M_\odot$ is 
probably not far from the true total mass. Our spectra 
taken in August and September 2005 showed a dark 
photospheric spot on the secondary component 
which became visible just after the secondary minimum (Fig.~\ref{fig5}). 
In addition to the main series of observations in 2005, 
V357~Peg had been shortly observed in 1997. These observations
do not show any indication of the photospheric spots. All radial
velocities are listed in Table~\ref{tab1}, but the 1997 data
were not used in orbital solution given in Table~\ref{tab2} to avoid
any influence of the possibly imprecisely-known or variable period. 
An orbit based on the 1997 data ($V_0 = -10.8$ km~s$^{-1}$, $K_1 =
93.8$ km~s$^{-1}$, $K_2 = 234.1$ km~s$^{-1}$, $T_0 = 2,450,748.7496(12)$
for the spectroscopic conjunction) is consistent with the 2005 results.

The trigonometric parallax of the system, $\pi = 6.00 \pm 1.56$ mas,
is too imprecise to determine reliably its absolute magnitude. 
Our estimate of the spectral type, F2V, agrees with both, 
the 2MASS color, $J-K = 0.178$, and the Tycho-2 color, $B-V =0.33$.

\subsection{V407~Peg}

V407~Peg (BD+14$\degr$5016; $V_{max}$ = 9.28) 
was found to be a variable during the Semi-Automatic Variability 
Search program at the Piwnice Observatory close
to Toru\'{n}, Poland \citep{maciej2002}. 
The authors presented a moderate-precision $BV$ photometry 
and the first ephemeris for the primary minima 
$HJD = 2\,452\,558.1703 + 0.636889 \times E$. The light curve
of V407~Peg was that of a contact binary. 

Later \citet{maciej2004} published 13 radial-velocity 
determinations based on spectra taken at the David Dunlap
Observatory and processed using the BF formalism. 
Unfortunately, about a half of the available spectra were taken close 
to the orbital conjunctions making the derived spectroscopic elements 
($V_0 = 22.1 \pm 5.9$ km~s$^{-1}$, 
$K_1 = 54.7 \pm 3.8$ km~s$^{-1}$ and 
$K_2 = 233.9 \pm 5.6$ km~s$^{-1}$) rather uncertain and
in fact very different from our orbit which is based
on 63 spectra. The most substantial difference is in the 8.7\%
larger sum of the semi-amplitudes resulting in the 29\% larger total 
(projected) mass. 

The star was not included in the {\it HIPPARCOS\/} mission. 
Its $J-K$ color in the 2MASS catalogue \citep{2MASS} 
is 0.181 corresponding to the F1 spectral type, 
which is consistent with the $(B-V) = 0.35$ determined 
by \citet{maciej2002}. Our spectral classification is F0V.

\subsection{V1123~Tau}

The variability of V1123~Tau (visual double WDS~03350+1743) was 
discovered during the {\it HIPPARCOS\/} satellite mission. 
In the {\it HIPPARCOS\/} Variable Stars Annex, the star
is classified as a $\beta$~Lyrae-type eclipsing binary with the 
ephemeris for the primary minimum: 
$HJD = 2\,458\,500.3570 + 0.399957 \times E$. 
The binary is accompanied by a 
fainter companion ($\rho = 4.3''$, $\theta = 136\degr$ 
and $\Delta V = 1.77$). The large error of the
{\it HIPPARCOS\/} trigonometric parallax,
$\pi = 6.82 \pm 3.03$ mas, suggests that it 
was most likely corrupted by the
presence of the visual companion.

\citet{ozdar2006} published the first ground-based photometry 
of the system. Based on the observed color of the system, 
$(B-V) = 0.684$, the authors estimated the spectral type to be G6V
which is much later than our direct classification of G0V. 
The rather red color and the wavelength-dependent depths of the
minima were very probably a result of the neglected 
contribution of the late-type companion which was almost 
certainly entering the diaphragm of the photoelectric photometer. 
The third light of the visual companion appears to have  
caused the amplitudes of the system to be color-dependent: 
while in the $U$ filter
the full amplitude of the light curve was 0.413 mag, it was
only 0.352 mag in the $R$ filter.

The light of the visual companion was partially 
entering the spectrograph slit during periods of the
poor seeing; it was however marginally visible 
in the extracted BFs with the relative contribution 
not larger than typically 0.02 -- 0.03. 
Its radial velocity, $V_3 = 29$ km~s$^{-1}$ was constant during 
our observing run and was fairly close to the center-of-mass 
velocity of eclipsing pair, $V_0 = 25.2 \pm 0.6$ km~s$^{-1}$. 

In addition to the recent observations reported here
(September 2005 -- March 2007),
V1123~Tau was also observed in 2002 using the older CCD chip 
(see previous papers of this series) and the 1800 lines/mm grating. The 
resulting radial velocities, listed in Table~\ref{tab1}, 
were not used in orbital solution given in Table~\ref{tab2} 
due to the long interval between the two datasets. The 
orbit based on the 2002 data
($V_0$ = 24.6 km~s$^{-1}$, $K_1$ = 69.3 km~s$^{-1}$, $K_2$ = 258.2 
km~s$^{-1}$, $T_0 = 2,452,558.4581(8)$ for the spectroscopic conjunction)
is consistent with the results from the new, better-defined data. 

The lower conjunction of the more massive component 
corresponds to the time of the secondary minimum
as given in the on-line database of ephemerides of 
eclipsing binary stars
(see http://www.as.ap.krakow.pl/ephem/ and Kreiner, 2004);
therefore V1123~Tau is W-type contact binary.

\subsection{V1128 Tau}

V1128 Tau (visual double WDS 03495+1255) is another {\it HIPPARCOS\/} discovery. 
Originally it was classified as a $\beta$~Lyrae-type eclipsing binary 
with the 0.3043732 days orbital period. The eclipsing
pair forms a relatively wide visual double with BD+12\degr511B, 
separated by 14 arcsec. The visual companion at the position 
angle $\theta = 196\degr$ was not entering our spectrograph slit 
which is permanently oriented
in the E-W direction. The high-precision photometry of 
\citet{tas2003} showed that the system is a
totally eclipsing contact binary, with the 
totality lasting about 16 minutes.
A subsequent light curve modeling led to $q_{ph} = 0.48$ and the
high orbital inclination angle of $i = 85$ degrees. The light curve 
asymmetry, with the maximum following the primary minimum being brighter,
was interpreted in terms of a cool spot on the cooler component.

Our spectroscopic
mass ratio, $q_{sp} = 0.534(6)$, is in a reasonable but not perfect
consistency with the photometric estimate. However, we could not 
see any evidence of dark photospheric spots on either of the components
in our BFs. 
The lower conjunction of the more massive component coincides with 
the deeper eclipse so that V1128~Tau is definitely a 
W-type contact binary. 

In addition to our recent observations reported here
(December 2005 -- March 2007), V1128~Tau had been shortly 
observed in 1997 and in 2003 using the old CCD detector and
the 1800 lines/mm grating. The resulting radial velocities, 
listed in Table~\ref{tab1}, were not used in orbital solution 
given in Table~\ref{tab2}. The corresponding orbit 
($V_0$ = $-$14.5 km~s$^{-1}$, $K_1$ = 127.1 km~s$^{-1}$, 
$K_2$ = 243.4 km~s$^{-1}$, $T_0 = 2,451,119.0927(5)$ for 
the spectroscopic conjunction) is
consistent with the results from the new data. 

The presence of the visual companion to the contact binary 
makes the {\it HIPPARCOS\/} parallax, $\pi = 1.71 \pm 6.27$ mas, 
rather uncertain (see \citet{priruc2006}). There is a
disparity between $J-K = 0.454$, as determined from 
the 2MASS survey, implying a spectral type later than G6V
and our spectral type estimate, F8V; this disparity 
is probably caused by the late-type visual companion.

\subsection{HH UMa}

The photometric variability of HH~UMa was discovered by the {\it HIPPARCOS\/} mission
where it was classified as a periodic variable with $P = 0.187747$ days.
On the basis of the color--period relation,
\citet{duer1997} concluded that it is very probably a 
genuine contact binary with twice the {\it HIPPARCOS\/} period. The ground-based
observations of \citet{prib2003} supported the contact binary nature of
HH~UMa; they also gave an improved ephemeris: 
$HJD = 2\,452\,368.3979 + 0.3754937 \times E$. 
Due to the partial eclipses, the mass ratio
could not be reliably determined and was estimated 
as $q_{ph} \approx 0.35 - 0.45$.

Our spectroscopy definitely shows a contact-binary nature 
of HH~UMa and gives $q_{sp} = 0.295(3)$. 
The low projected total mass, $(M_1 + M_2) \sin^3 i = 0.826 M_\odot$, 
together with low amplitude of the light curve, 0.17 mag, support the low 
inclination angle, as found in the photometric analysis.

The trigonometric parallax of the system, $\pi = 4.50 \pm 1.78$ 
mas, is too imprecise to draw any conclusions on its absolute magnitude. 
The 2MASS color of the system, $J-K = 0.572$, agrees fairly 
well with our estimate of the spectral type,
F5V. The Tycho-2 $B-V = 0.50$ corresponds to the F7V spectral type.

\subsection{PY Vir}

PY~Vir (GSC 4961-667) was found on Stardial images to be a
variable of the W~UMa type \citep{wils2003}. 
The system is also known as an X-ray source
(1RXS J131032.4-040934). No photometric or 
spectroscopic analysis of PY~Vir
has been published yet. A minimum observed photometrically 
in parallel with the spectroscopic observations
($HJD = 2\,554\,201.7944$) was used for 
a preliminary improvement of the ephemeris used for phasing of
our spectroscopy.

The broadening functions of PY~Vir based on the 2007 observations 
show the presence of a third component with a constant
radial velocity, $V_3 = -32.9 \pm 4.1$ km~s$^{-1}$; this velocity
is rather different from the 
systemic velocity of the contact binary, $V_0 = -14.66$ km~s$^{-1}$.
A few additional, low quality spectra of PY~Vir taken 
in poor-weather conditions on February 19, 2008
(not listed in Table~\ref{tab1}) give $V_3 = -23.3 \pm 3.7$ km~s$^{-1}$
(at the mean HJD = 2\,454\,515.883) so that we have an indication of
a change in $V_3$. The velocities of the close binary from these
additional observations are not precise enough to see if the motion of the 
third component is reflected in systemic radial 
velocity of the eclipsing pair. The third component may be a 
binary itself; otherwise, the velocity changes may
result from its low mass and a relatively fast orbital motion in a
tight triple system. We consider the latter
possibility as a relatively probable and more exciting one,
but its verification would require observations over several seasons.

PY~Vir has not been known to be a multiple system; it is also not listed
in the WDS catalogue. A lunar occultation of the system in July 1984 did not
reveal presence of any visual companion \citep{evans1985}, hence the separation
of components is probably very small. Moreover, the light contribution of the 
third component is fairly small, only about $L_3/(L_1 + L_2) = 0.08$, at the 
maximum light of the eclipsing pair.

The 2MASS color of the system, $J-K = 0.572$, indicates the
K2V spectral type, which is consistent with our spectral classification.
The orbital period of the system, 0.311 days, is rather long for such
a late spectral type (it would be more consistent with late F)
so that the system may be close but detached. The shape of the
broadening function (see Figure~\ref{fig4}) with a gap between the
two peaks is not inconsistent with this possibility.
The Tycho-2 $(B-V)$ color is rather uncertain to 
draw any conclusions, $B-V = 0.80 \pm 0.06$.

\section{SUMMARY}
\label{summary}

With the new ten short-period binaries, this paper brings the
number of the systems studied at the David Dunlap Observatory to 120.
Almost all systems of this group have been rather
neglected and little has been known about them. 

The highlights of this series are: 
(1)~the triple systems BD+7$\degr$3142 and PY~Vir,
both with short-period, late spectral-type (K2V)
binaries; (2)~both of the above are interesting: BD+7$\degr$3142
is a high space-velocity system while PY~Vir
maybe a very close, but detached binary; 
(3)~the spotted contact binary, V1128~Tau,
with a large spot on the secondary component, (4)~V1191~Cyg,
the unusual, W-type contact binary with the small mass ratio of 0.107 and
a relatively short-period, observed during the rapid mass-transfer stage.
V1191~Cyg has an unexpectedly short orbital period for its 
F6V spectral type implying smaller components than for
typical solar-composition, solar-age contact binaries;
this may be an indication of its old
population characteristics, similar to those of binaries in
globular clusters \citep{rci2000}.
None of the systems had been observed spectroscopically before 
(except for the measurements of the rotational 
velocity of the EG~Cep primary component \citep{etzel1993}). 

\acknowledgements

We express our thanks to  Michal Siwak, George Conidis, Wojtek Pych, 
and Mel Blake for their contribution to
the observations and to Wojtek Pych for his cosmic-ray removal code.

Support from the Natural Sciences and Engineering Council of Canada
to SMR and SWM and from the Polish Science Committee (KBN/MNiSW
grants PO3D~006~22, PO3D~003~24, PO3D~025~29) is 
acknowledged with gratitude. The travel of TP to
Canada has been supported by a Slovak Academy of Sciences
VEGA grant 2/7010/7.

The research made use of the SIMBAD database, operated at the CDS,
Strasbourg, France and accessible through the Canadian
Astronomy Data Centre, which is operated by the Herzberg Institute of
Astrophysics, National Research Council of Canada.
This research made also use of the Washington Double Star (WDS)
Catalog maintained at the U.S. Naval Observatory.

\clearpage

\noindent
Captions to figures:

\bigskip

\figcaption[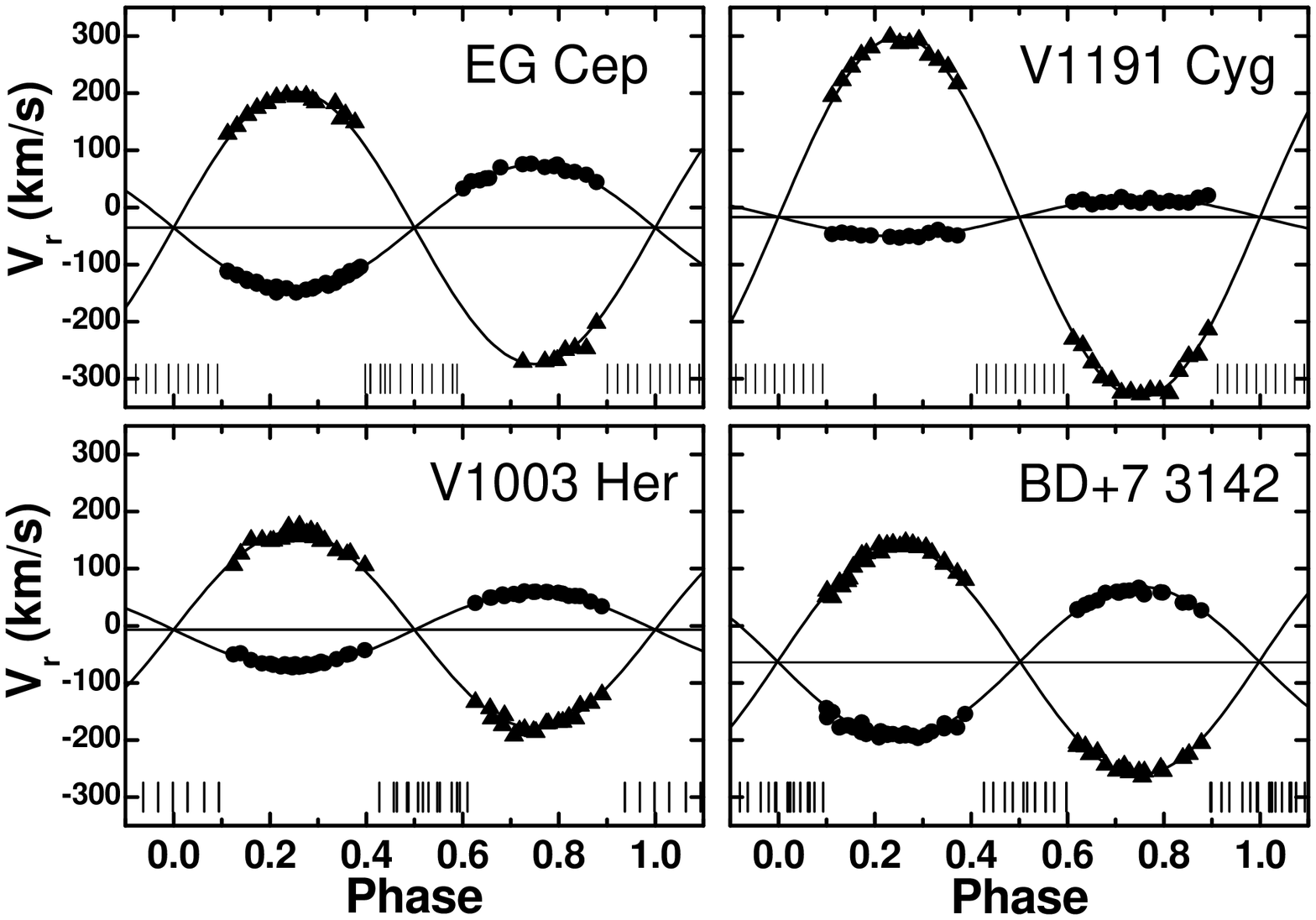] {\label{fig1} Radial velocities of the
systems EG~Cep, V1191~Cyg, V1003~Her and BD+7$\degr$3142 (in Hercules)
are plotted in individual panels versus the orbital phases. The 
lines give the respective circular-orbit (sine-curve) fits to the RV's.
EG~Cep is very probably a detached or semi-detached binary,
V1003~Her is close binary of uncertain type while 
V1191~Cyg and BD+7$\degr$3142
are contact binaries. BD+7$\degr$3142 is member of triple system
with a relatively bright third component. 
The circles and triangles in this and
the next two figures correspond to components with velocities
$V_1$ and $V_2$, as listed in Table~\ref{tab1}, respectively. The
component eclipsed at the minimum corresponding to $T_0$ (as given
in Table~\ref{tab2}) is the one which shows negative velocities for
the phase interval $0.0 - 0.5$ and which is the more massive one.
Short marks in the lower parts of the panels show phases of available
observations which were not used in the solutions because of the
excessive spectral line blending. }

\figcaption[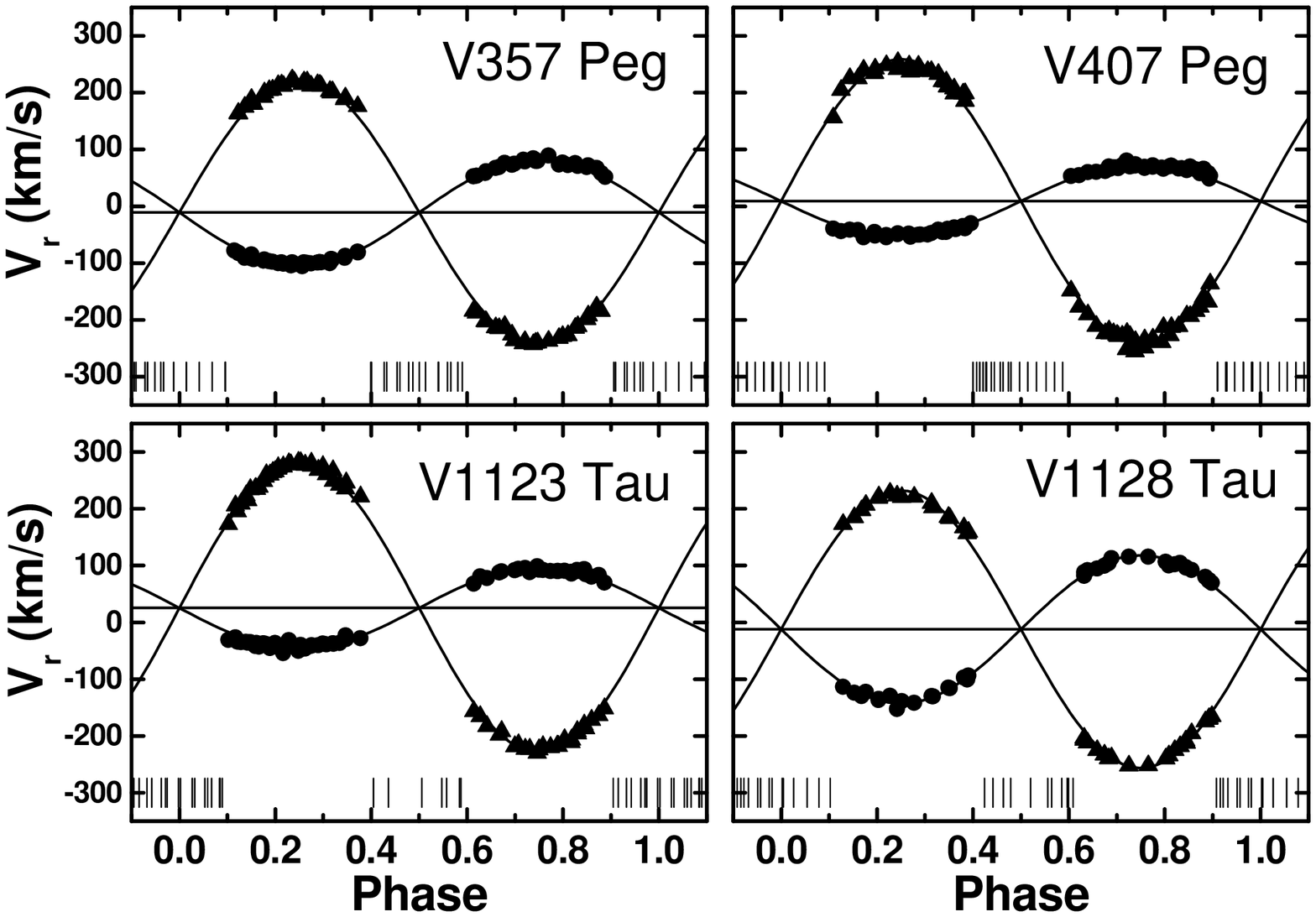] {\label{fig2} The same as for
Figure~\ref{fig1}, but for V357~Peg, V407~Peg, V1123~Tau, and V1128~Tau.
All four systems are contact binaries. V1123~Tau and V1128~Tau are
members of relatively wide visual binaries.}

\figcaption[rvs9-10.ps] {\label{fig3} The same as for
Figures~\ref{fig1} and \ref{fig2}, for the systems
HH~UMa, and PY~Vir. Both are typical contact binaries.}

\figcaption[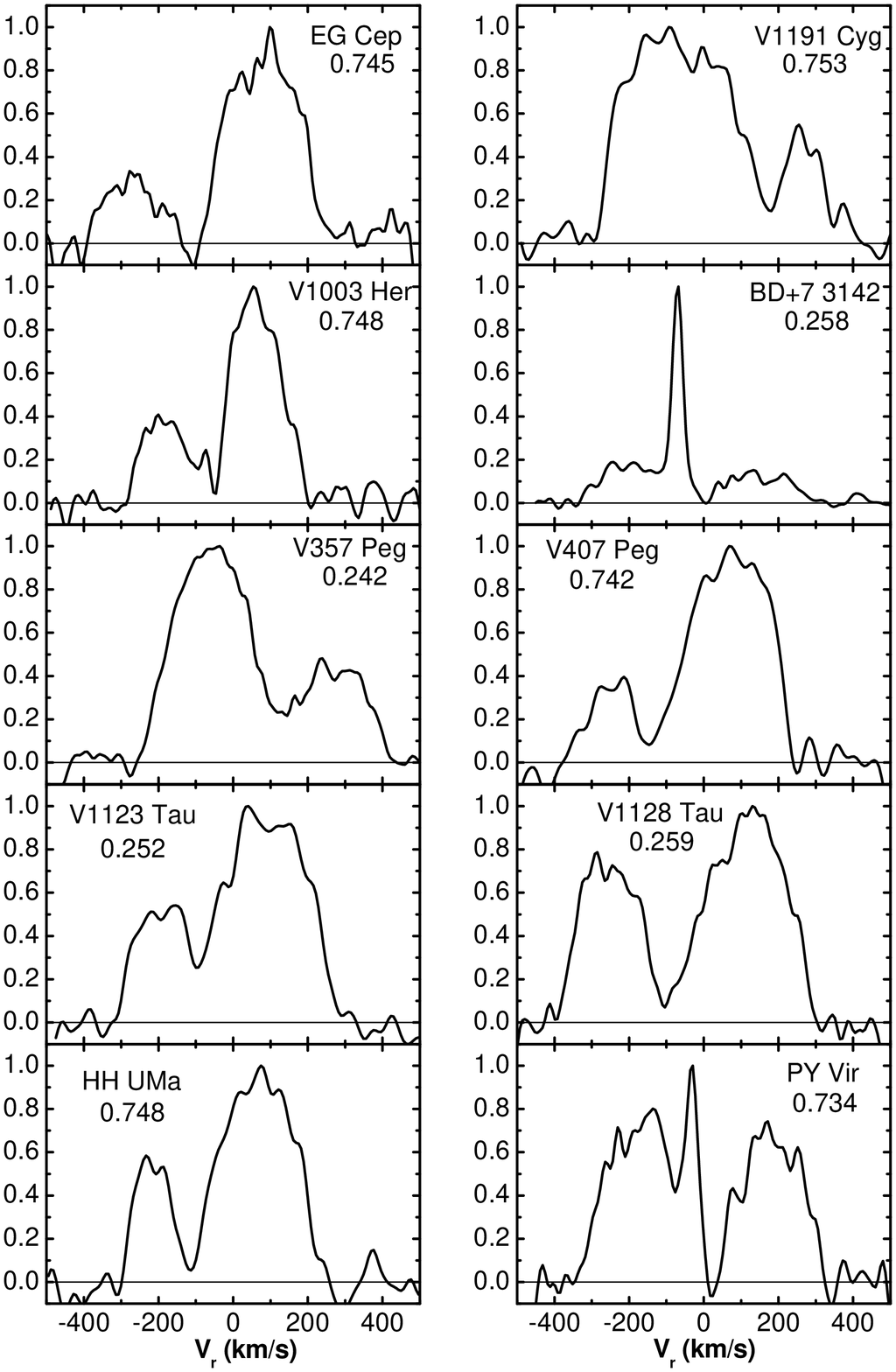] {\label{fig4}
The broadening functions (BFs) for all ten systems of this
group, selected for phases close to 0.25 or 0.75.
The phases are marked by numbers in individual panels.
The third star features in the BFs of the contact
binaries BD+7$\degr$3142 and PY~Vir are strong and clearly visible.
All panels have the same horizontal range, $-500$ to 
$+500$ km~s$^{-1}$.}

\figcaption[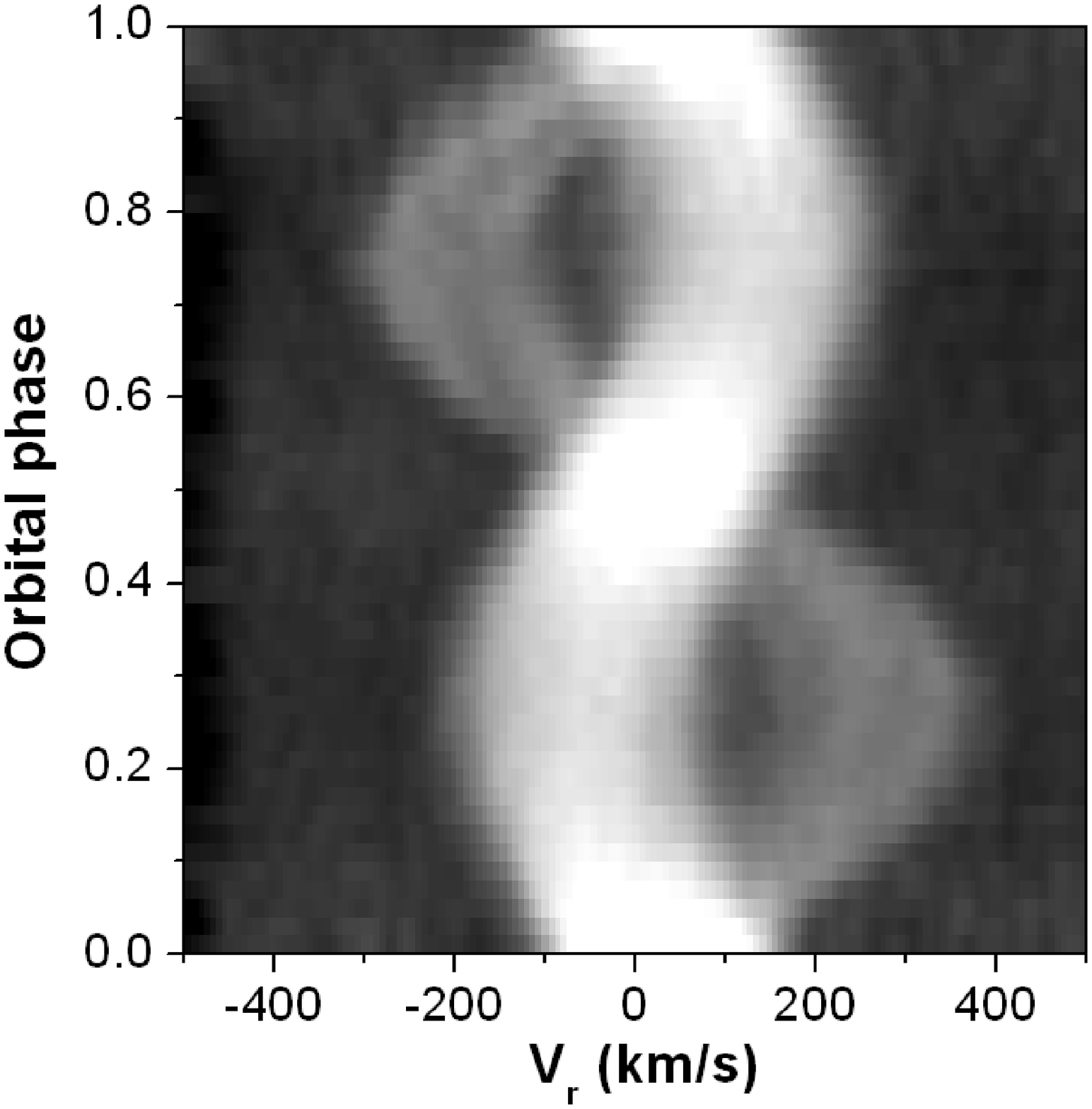] {\label{fig5}
The broadening functions (BFs) of V357~Peg determined from
observations between August 25 and September 6, 2005. The
original BFs were binned into 0.02 phase intervals and smoothed
by convolution with a Gaussian profile ($\sigma = 0.02$ in phase).
The dark feature drifting through the secondary component
profile is very probably a large photospheric spot. Its signature
appears to be variable, especially around the phase 0.75 implying
that the spot
may have changed its position during the two weeks of our observations.}

\clearpage

\addtocounter{figure}{-5}

\begin{figure} 
\epsscale{0.85}
\plotone{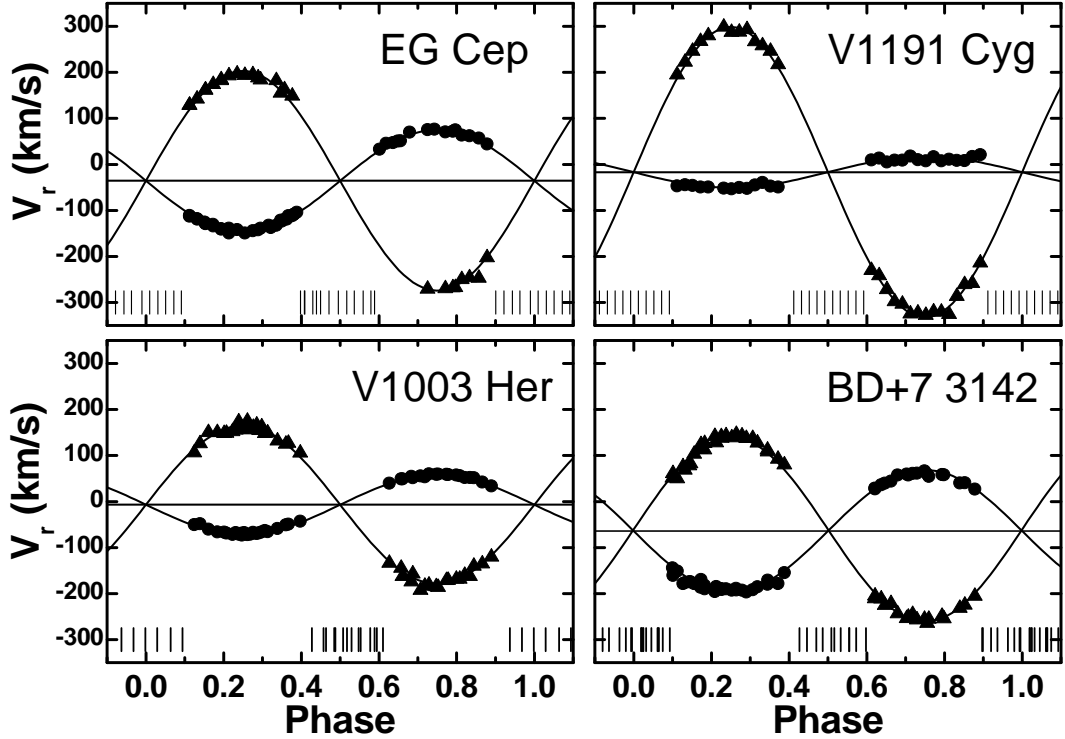}
\caption{Radial velocities of the
systems EG~Cep, V1191~Cyg, V1003~Her and BD+7$\degr$3142 (in Hercules)
are plotted in individual panels versus the orbital phases. The 
lines give the respective circular-orbit (sine-curve) fits to the RV's.
EG~Cep is very probably a detached or semi-detached binary,
V1003~Her is close binary of uncertain type while 
V1191~Cyg and BD+7$\degr$3142
are contact binaries. BD+7$\degr$3142 is member of triple system
with a relatively bright third component. 
The circles and triangles in this and
the next two figures correspond to components with velocities
$V_1$ and $V_2$, as listed in Table~\ref{tab1}, respectively. The
component eclipsed at the minimum corresponding to $T_0$ (as given
in Table~\ref{tab2}) is the one which shows negative velocities for
the phase interval $0.0 - 0.5$ and which is the more massive one.
Short marks in the lower parts of the panels show phases of available
observations which were not used in the solutions because of the
excessive spectral line blending.}
\end{figure}

\begin{figure} 
\epsscale{0.85}
\plotone{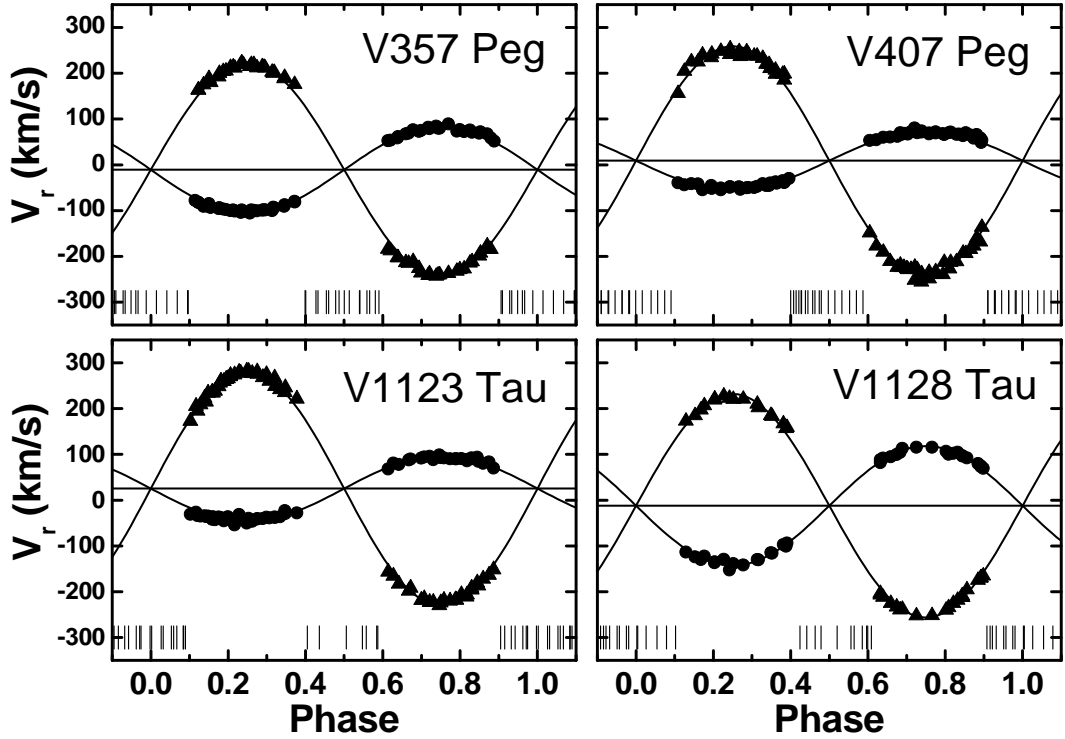}
\caption{The same as for
Figure~\ref{fig1}, but for V357~Peg, V407~Peg, V1123~Tau, and V1128~Tau.
All four systems are contact binaries. V1123~Tau and V1128~Tau are
members of relatively wide visual binaries.}
\end{figure}

\begin{figure} 
\epsscale{0.85}
\plotone{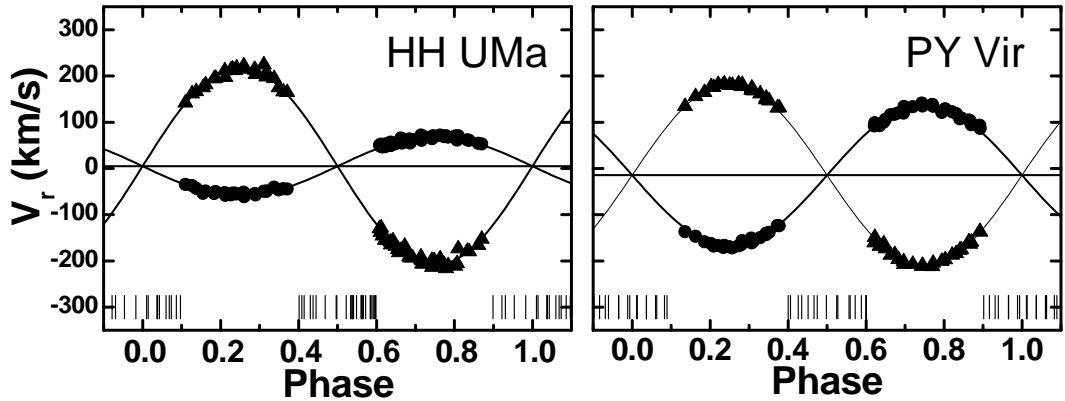}
\caption{The same as for
Figures~\ref{fig1} and \ref{fig2}, for the systems
HH~UMa, and PY~Vir. Both are typical contact binaries.}
\end{figure}

\begin{figure} 
\epsscale{0.75}
\plotone{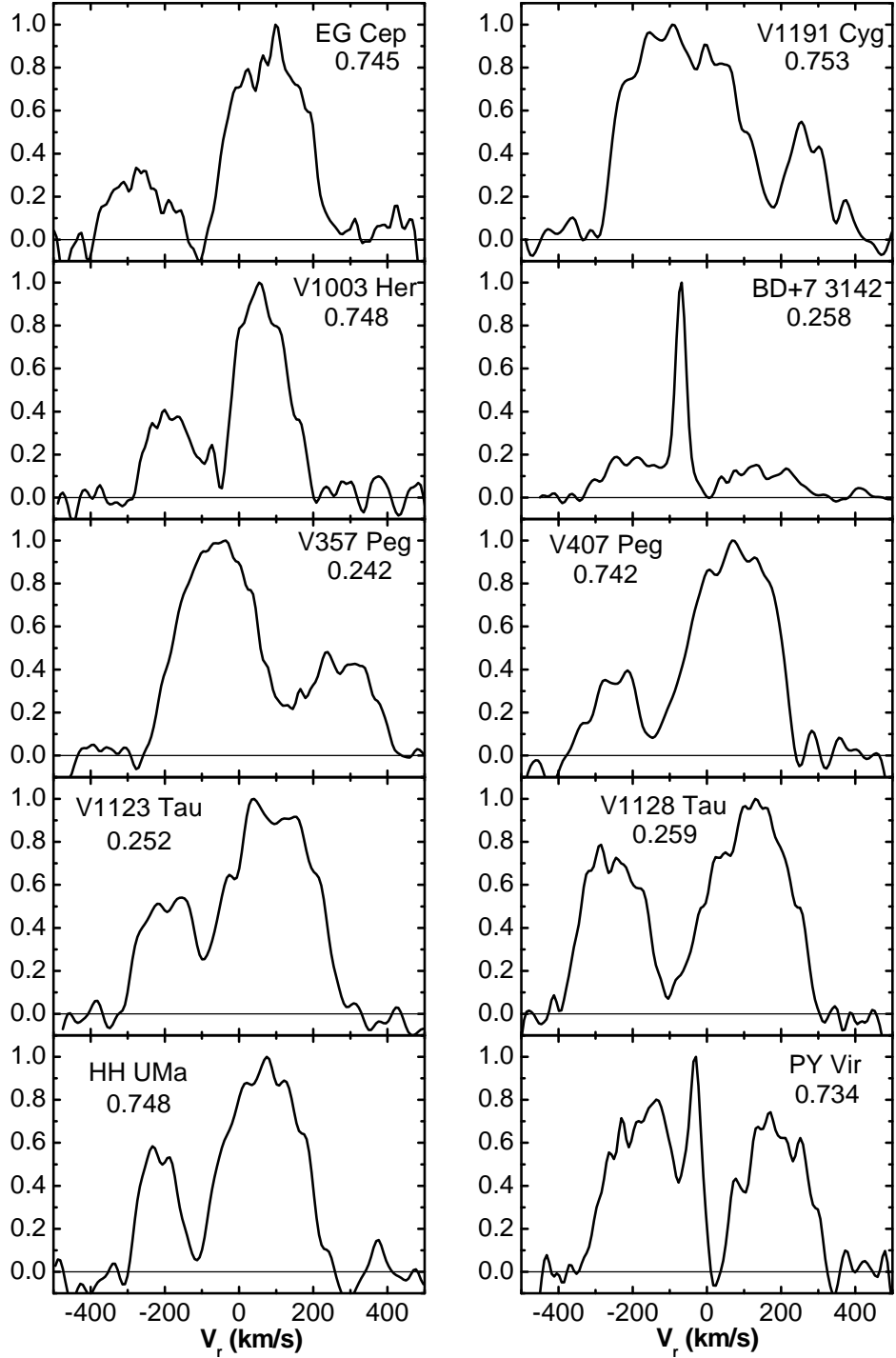}
\caption{The broadening functions (BFs) for all ten systems of this
group, selected for phases close to 0.25 or 0.75.
The phases are marked by numbers in individual panels.
The third star features in the BFs of the contact
binaries BD+7$\degr$3142 and PY~Vir are strong and clearly visible.
All panels have the same horizontal range, $-500$ to 
$+500$ km~s$^{-1}$.}
\end{figure}

\begin{figure} 
\epsscale{0.85}
\plotone{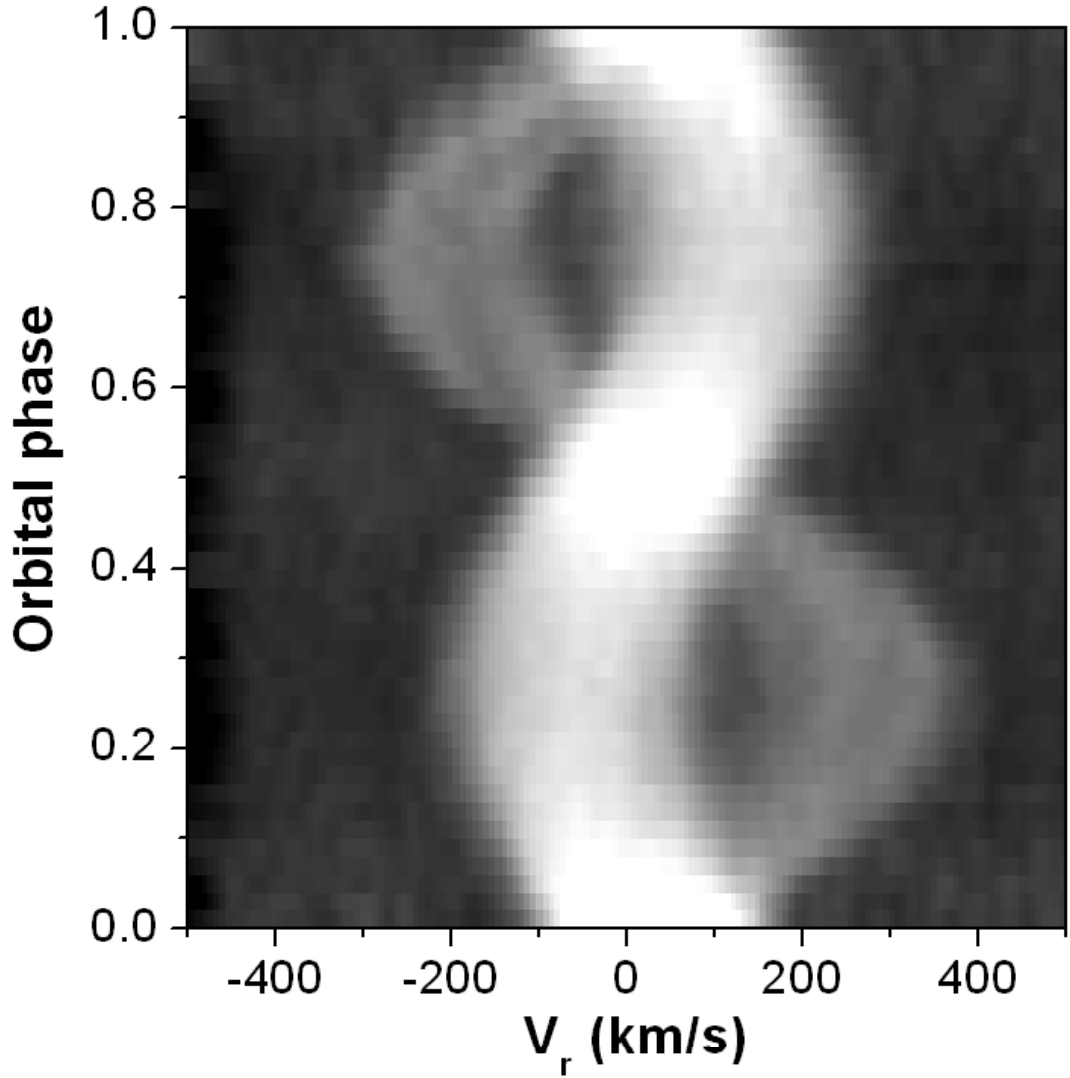}
\caption{The broadening functions (BFs) of V357~Peg determined from
observations between August 25 and September 6, 2005. The
original BFs were binned into 0.02 phase intervals and smoothed
by convolution with a Gaussian profile ($\sigma = 0.02$ in phase).
The dark feature drifting through the secondary component
profile is very probably a large photospheric spot. Its signature
appears to be variable, especially around the phase 0.75 implying
that the spot
may have changed its position during the two weeks of our observations.}

\end{figure}

\begin{deluxetable}{lrrrrr}

\tabletypesize{\footnotesize}

\tablewidth{0pt}
\tablenum{1}

\tablecaption{DDO radial velocity observations (the full
table is available only in electronic form)
\label{tab1}}

\tablehead{
\colhead{HJD--2,400,000}  &
\colhead{~$V_1$}          &
\colhead{~~$W_1$}         &
\colhead{~$V_2$}          &
\colhead{~~$W_2$}         &
\colhead{Phase}          \\
                          &
\colhead{[km s$^{-1}$]}   &
                          &
\colhead{[km s$^{-1}$]}   &
                          & \\
}
\startdata
 54251.6402 & $-139.75$& 1.00 & 190.61  & 1.00 & 0.2919 \\
 54251.6582 & $-134.70$& 1.00 & 183.19  & 0.00 & 0.3250 \\
 54251.6736 & $-120.90$& 1.00 & 162.99  & 0.00 & 0.3533 \\
 54251.6902 & $-106.37$& 1.00 & 167.77  & 0.00 & 0.3837 \\
 54251.7059 &     0.00 & 0.00 &   0.00  & 0.00 & 0.4126 \\
 54251.7222 &     0.00 & 0.00 &   0.00  & 0.00 & 0.4425 \\
 54251.8037 &     0.00 & 0.00 &   0.00  & 0.00 & 0.5921 \\
 54251.8199 &    48.14 & 1.00 &$-211.99$& 0.00 & 0.6218 \\
 54251.8363 &    53.42 & 1.00 &$-258.35$& 0.00 & 0.6520 \\
 54251.8527 &    72.41 & 1.00 &$-254.16$& 0.00 & 0.6820 \\
\enddata

\tablecomments{The table gives the radial velocities used  
in the paper. The first 10 rows of the table for the 
first program star, EG~Cep, are shown. Observations leading to 
entirely inseparable broadening function peaks are given a zero weight; 
these observations may be eventually used in more extensive modeling of 
broadening functions. The zero weight 
was also assigned to observations of marginally 
visible peaks of the secondary component (mainly in the case of EG~Cep).
The RV's designated as $V_1$ correspond to
the more massive component; it was always the component eclipsed
during the minimum at the epoch $T_0$ (this does not always correspond
to the deeper minimum and photometric phase 0.0). The phases listed
in the last column
correspond to $T_0$ and periods given in Table~2. For V1191~Cyg, where
we used phase-smoothed BFs, the heliocentric Julian dates are omitted.}

\end{deluxetable}

\begin{deluxetable}{lccrrrccc}

\tabletypesize{\scriptsize}

\pagestyle{empty}
\tablecolumns{9}

\tablewidth{0pt}

\tablenum{2}
\tablecaption{Spectroscopic orbital elements \label{tab2}}
\tablehead{
   \colhead{Name} &                
   \colhead{Type} &                
   \colhead{Other names} &         
   \colhead{$V_0$~~~} &            
   \colhead{$K_1$~~~} &            
   \colhead{$\epsilon_1$~} &       
   \colhead{T$_0$ -- 2,400,000} &  
   \colhead{P (days)} &            
   \colhead{$q$}          \\       
   \colhead{}     &                
   \colhead{Sp.~type}    &         
   \colhead{}      &               
   \colhead{} &                    
   \colhead{$K_2$~~~} &            
   \colhead{$\epsilon_2$~} &       
   \colhead{$(O-C)$(d)~[E]} &      
   \colhead{$(M_1+M_2) \sin^3i$} & 
   \colhead{}                      
}
\startdata
EG~Cep     & EB                  & HD194089    & $-35.61$(0.74)  & 110.67(1.02) &
      3.41 & 54304.3114(8)       & 0.54462228  & 0.464(5)        \\ 
           & A7V                 & BD+76$\degr$790   &                 & 238.72(1.30) &
      8.67 & $+0.0010$~[+3,312.0]& 2.407(27)   &                 \\[1mm]

V1191~Cyg  & EW(W)               &             & $-16.82$(0.94)  &  33.68(1.52) &
      6.28 & 52500.4104(8)       & 0.3133867   & 0.107(5)        \\ 
           & F6V                 &             &                 & 315.52(1.52) &
      8.16 & $+0.0030$~[+0.5]    & 1.383(22)   &                 \\[1mm]

V1003~Her  & EW?                 & HD343341    & $-6.92$(0.61)   &  64.07(0.94) &
      2.92 & 54286.5955(9)       & 0.493322    & 0.373(6)        \\ 
           & A7V                 & BD+21$\degr$3589  &                 & 171.91(0.94) &
      7.66 & $-0.0082$~[+11,729] & 0.672(9)    &                 \\[1mm]

BD+07$\degr$3142 & EW?                 &             & $-64.08$(0.79)  & 132.56(1.27) &
      7.82 & 54212.8136(4)       & 0.2752770   & 0.662(8)        \\ 
           & K2V                 &             &                 & 200.16(1.44) &
      8.83 & $-0.0018$~[+87]     & 1.050(14)   &                 \\[1mm]

V357~Peg   & EW(A)               & HD222994    & $-10.84$(0.54)  &  93.78(0.86) &
      4.07 & 53730.6697(6)       & 0.5784514   & 0.401(4)        \\ 
           & F2V                 & BD+24$\degr$4828  &                 & 234.08(0.87) &
      7.19 & $+0.0024$~[+2,127]  & 2.112(18)   &                 \\[1mm]

V407~Peg   & EW(A)               & BD+14$\degr$5016  & $9.06$(0.70)    &  63.92(1.12) &
      4.59 & 54049.7666(8)       & 0.636889    & 0.256(6)        \\ 
           & F0V                 &             &                 & 250.02(1.16) &
     10.22 & $+0.0023$~[+2,342]  & 2.042(25)   &                 \\[1mm]

V1123~Tau  & EW(W)               & BD+17$\degr$579   & $25.32$(0.54)   &  71.08(0.84) &
      5.42 & 53791.0975(4)       & 0.3999496(8)& 0.279(4)        \\ 
           & G0V                 & HIP 16706   &                 & 254.71(0.84) &
      6.41 & $+0.0004$~[+3,380.5]& 1.433(13)   &                 \\[1mm]

V1128~Tau  & EW(W)               & BD+12$\degr$511   & $-12.27$(0.76)  & 130.48(1.27) &
      5.75 & 53864.7026(4)       & 0.3053707(7)& 0.534(6)        \\ 
           & F8V                 & HIP 17878   &                 & 244.19(1.28) &
      7.68 & $-0.0007$~[+4,468.5]& 1.664(18)   &                 \\[1mm]

HH~UMa     & EW(A)               & BD+36$\degr$2149  & $4.92$(0.44)    &  63.06(0.69) &
      3.74 & 54,070.1229(4)      & 0.3754889(9)& 0.295(3)        \\ 
           & F5V                 & HIP~54165   &                 & 213.84(0.75) &
      6.17 & $-0.0017$~[+4,181]  & 0.826(7)    &                 \\[1mm]

PY~Vir     & EW(W)               &             & $-14.51$(0.45)  & 152.78(0.77) &
      4.92 & 54,193.2347(3)      & 0.311251    & 0.773(5)        \\ 
           & K1/2V               & BD-03$\degr$3419  &                 & 197.58(0.77) &
      4.65 & $-0.0149$~[+7,364.5]& 1.387(10)   &                 \\[1mm]

\enddata
\tablecomments{The spectral types given in column 2 
relate to the combined spectral
type of all components in a system; they are given 
in parentheses if taken from the
literature, otherwise are new. The convention of naming 
the binary components in the  table is that the 
more massive star is marked by the subscript ``1'', so that the
mass ratio is defined to be always $q \le 1$. 
The standard errors of the circular solutions in 
the table are expressed in units of last decimal 
places quoted; they  are given in parentheses 
after each value. The center-of-mass velocities ($V_0$),
the velocity amplitudes ($K_i$) and the standard 
unit-weight errors of the solutions
($\epsilon$) are all expressed in km~s$^{-1}$. 
The spectroscopically determined  moments of primary 
or secondary minima are given by $T_0$; the corresponding
$(O-C)$ deviations (in days) have been 
calculated from the available prediction on
primary minimum, as given in the text, using 
the assumed periods and the number of
epochs given by [E]. 
The values of $(M_1+M_2)\sin^3i$ are in the solar mass units.\\
 Ephemerides ($HJD_{min}$ -- 2,400,000 + period in days) 
used for the computation of the $(O-C)$ residuals:\\
 EG~Cep:    52500.5214 + 0.54462228 \\ 
 V1191~Cyg: 52500.2507 + 0.3133874 \\  
 V1003~Her: 48500.43   + 0.493322 \\   
 BD+7$\degr$3142: 54188.8663 + 0.275277 \\   
 V357~Peg:  52500.3022 + 0.5784509 \\  
 V407~Peg:  52558.1703 + 0.636889 \\   
 V1123~Tau: 52500.2670 + 0.3999474 \\  
 V1128~Tau: 52500.1463 + 0.3053725 \\  
 HH~UMa:    52500.1967 + 0.3754910 \\  
 PY~Vir:    51901.0416 + 0.311251    
}
\end{deluxetable}

\end{document}